\documentclass[runningheads]{llncs}
\usepackage{stackengine}
\newcommand\barbelow[1]{\stackunder[1.2pt]{$#1$}{\rule{.8ex}{.075ex}}}
\usepackage{epsfig}
\usepackage{graphicx}
\usepackage{amsmath}
\usepackage{amssymb}
\usepackage{lipsum}
\usepackage{enumitem}
\usepackage[belowskip=-12pt,aboveskip=-6pt]{caption}
\usepackage[pagebackref=true,breaklinks=true,colorlinks,bookmarks=false]{hyperref}
\usepackage{makecell}
\usepackage{array}
\usepackage{cite}
\usepackage[toc, page]{appendix}

\usepackage{array, makecell}
\usepackage{boldline}
\usepackage[x11names, table]{xcolor}
\usepackage{colortbl}
\begin{document}
\title{Spatially-varying Regularization with Conditional Transformer for Unsupervised Image Registration}
\titlerunning{Spatially-varying Regularization for Unsupervised Image Registration}
\author{Junyu Chen \inst{1} \and Yihao Liu \inst{2} \and Yufan He \inst{3} \and Yong Du \inst{1} \\\email{\{jchen245,yliu236,duyong\}@jhmi.edu;yufanh@nvidia.com}}
%
%\author{Anonymous MICCAI 2023 submission}
\authorrunning{J. Chen \emph{et al.}}
% If there are more than two authors, '\emph{et al.}' is used.
%
\institute{Russell H. Morgan Department of Radiology and Radiological Science,\\ Johns Hopkins Medical Institutes, Baltimore, MD, USA \and Department of Electrical and Computer Engineering,\\ Johns Hopkins University, Baltimore, MD, USA \and NVIDIA Corporation, Bethesda, MD, USA}
\maketitle              % typeset the header of the contribution
\begin{abstract}
In the past, optimization-based registration models have used spatially-varying regularization to account for deformation variations in different image regions. However, deep learning-based registration models have mostly relied on spatially-invariant regularization. Here, we introduce an end-to-end framework that uses neural networks to learn a spatially-varying deformation regularizer directly from data. The hyperparameter of the proposed regularizer is conditioned into the network, enabling easy tuning of the regularization strength. The proposed method is built upon a Transformer-based model, but it can be readily adapted to any network architecture. We thoroughly evaluated the proposed approach using publicly available datasets and observed a significant performance improvement while maintaining smooth deformation. The source code of this work will be made available after publication.
\keywords{Image Registration \and Spatially Varying Regularization.}
\end{abstract}

\section{Introduction}
In recent years, the advancement of deep neural networks (DNNs) and their success in processing image data has sparked increased interest in developing DNN-based methods for medical image registration.
Deep learning-based registration methods train a DNN on an image dataset to optimize a global objective function, usually in the form of a similarity measure, with a regularizer added to enforce the spatial smoothness of the deformation.
The strength of regularization is often controlled by a fixed and spatially-invariant hyperparameter in most existing methods~\cite{balakrishnan2019voxelmorph, dalca2019unsupervised, de2017end, chen2021vitvnet,chen2022transmorph,liu2022coordinate, kim2021cyclemorph}. While some methods have been proposed to condition the hyperparameter into the network architecture~\cite{mok2021conditional, hoopes2021hypermorph}, the regularization remains spatially invariant, meaning the same regularization strength is applied \textit{everywhere} in the deformation. However, such a regularization may not be optimal in many situations, as it does not account for the variations of deformation that may be necessary for different regions of the image~\cite{niethammer2019metric}. One example of this can be seen in the registration of brain scans, where the ventricles in the brain can vary in size among different patients, leading to different scales of deformation in the ventricles compared to other parts of the brain~\cite{niethammer2019metric}. Similarly, when registering inhale-to-exhale images of the lung, larger deformations of the lung than those of the surrounding tissue are anticipated~\cite{shen2019region}.

Efforts have been made to develop regularizers that vary in space, including optimization-based registration schemes (\emph{e.g.}\cite{pace2013locally, risser2013piecewise, stefanescu2004grid, vialard2014spatially, gerig2014spatially, kabus2006variational}) and deep learning-based methods that learn spatially varying regularization from data (\emph{e.g.}\cite{niethammer2019metric, shen2019region}). In \cite{niethammer2019metric}, Niethammer \emph{et al.} proposed using DNNs to predict locally adaptive weights for multi-Gaussian kernels, but their method requires pre-setting the variance of the kernels and is not effective for end-to-end registration networks.
Shen \emph{et al.}~\cite{shen2019region} expanded upon the work of \cite{niethammer2019metric} by introducing an end-to-end training scheme for learning a spatially-varying regularizer and the initial momentum in a spatio-temporal velocity setting.
This method has the ability to track the deformation of regions and estimate a unique regularizer for each time point. However, the implementation of this method is not straightforward, and it cannot be seamlessly integrated with existing DNN-based methods.

In this paper, a novel DNN method is introduced for end-to-end learning of a spatially-varying regularizer from data. The main contributions of this work can be summarized as follows: \textbf{1)} We proposed a weighted diffusion regularizer that applies spatially varying regularization to the deformation and uses a novel log loss to ensure overall smoothness. Although the proposed method was demonstrated in the form of a diffusion regularizer, it is readily adaptable to other types of regularizers such as anisotropic diffusion and bending energy. \textbf{2)} We introduced a hyperparameter conditioning method to a Transformer-based network and conditioned the regularization hyperparameter into the network architecture, enabling the ability to capture varying levels of spatially varying regularization through a single training process and control the smoothness of deformation during inference. \textbf{3)} The proposed method is straightforward to implement and can be adapted to various existing unsupervised DNN models without the need for significant changes to the script. \textbf{4)} The proposed method was thoroughly evaluated using publicly available datasets and showed improvement in performance in comparison to the baseline models while maintaining low levels of deformation irregularities. 

\section{Backgrounds and Related Works}
\label{sec:background}
\noindent{\textbf{Diffusion Regularizer. }}
In learning-based image registration, the diffusion regularizer is often used to impose deformation smoothness~\cite{balakrishnan2019voxelmorph, dalca2019unsupervised, kim2021cyclemorph, chen2022transmorph, chen2021vitvnet, liu2022coordinate}. It is expressed as:
\begin{equation}
\label{eqn:diff}
    \mathcal{R}(\phi) = \sum_{\mathbf{p}\in\Omega}\Vert\nabla \pmb{u}(\mathbf{p})\Vert^2,
\end{equation}
where $\pmb{u}$ denotes the displacement field, $\mathbf{p}$ is the voxel location, and $\phi\subset\mathbb{R}^3$ represents the 3D spatial domain. The operator, $\nabla$, computes the spatial gradients, which can be approximated using finite differences. As shown in \cite{dalca2019unsupervised, dalca2019learning}, this regularizer can be derived from the maximum posterior estimation of the variable $\pmb{u}$ by assuming the prior distribution over $\pmb{u}$ to be a multivariate Normal distribution with mean $\pmb{\mu}=\mathbf{0}$ and covariance $\mathbf{\Sigma}$: $p(\pmb{u}) \propto \mathcal{N}(\pmb{u};\barbelow{\mathbf{0}},\mathbf{\Sigma})$.
Let $\mathbf{\Sigma}^{-1}=\mathbf{\Lambda}_{\pmb{u}}=\lambda \mathbf{L}$, where $\mathbf{L}$ is the Laplacian of a neighborhood graph defined on the image grid, and $\lambda$ is a parameter that controls the scale of the displacement, the logarithm of this prior can be simplified to Eqn. \ref{eqn:diff}.:
\begin{equation}
   \log p(\pmb{u}) \propto -\pmb{u}^\intercal\mathbf{\Lambda}_{\pmb{u}}\pmb{u}=-\lambda\pmb{u}^\intercal\mathbf{L}\pmb{u}=-\lambda\sum_{\mathbf{p}\in\Omega}\Vert\nabla \pmb{u}(\mathbf{p})\Vert^2,
\end{equation}
by noting the fact that $\log\vert\mathbf{\Lambda}_{\pmb{u}}\vert$ is a constant. The parameter $\lambda$ regulates the smoothness of the deformation by varying the covariance of the Normal distribution. When $\lambda$ is increased, the covariance $\mathbf{\Sigma}$ decreases, resulting in greater similarity among neighboring displacements. This can also be understood as applying a larger variance Gaussian kernel to the displacement field, leading to a smoother deformation.

\noindent{\textbf{Learning Spatially-varying Regularizer. }}
In \cite{niethammer2019metric}, a method for spatially varying regularization through metric learning was proposed. The approach involves learning locally adaptive weights for multiple Gaussian kernels with different standard deviations, in the form: $\sum_{i=0}^{N-1} w_i(\mathbf{p}) G_i$.
Here, $w_i(\mathbf{p})$ represents the weight of the $i^{th}$ Gaussian kernel $G$ at the voxel location $\mathbf{p}$, and they satisfy $\sum_{i=0}^{N-1} w_i(\mathbf{p})=1$.
A neural network $f$, with parameters $\theta$, predicts these adaptive weights using inputs of the image pair and the initial momentum $m_0$ generated by a traditional registration method.
This can be represented as $[w_0,\ldots,w_{N-1}]=f_\theta(I_m, I_f, m_0)$. Then, the deformation is smoothed by convolving the weighted multi-Gaussian kernel with $m_0$.
%\begin{equation}
%\sum_{i=0}^{N-1} \sqrt{w_i(\mathbf{p})} \int_\mathbf{q}G_i(\mathbf{p}-\mathbf{q})\sqrt{w_i(\mathbf{q})}m_0(\mathbf{q})d\mathbf{q}.
%\end{equation}
During network training, two loss functions were placed over $w$'s to promote smoothness in the momentum and adaptive weights.
These loss functions included an optimal mass transport (OMT) loss and a total variational (TV) loss. The OMT loss imposes the network to prioritize the use of the Gaussian kernel with the largest variance, while the TV loss encourages weight changes coinciding with image edges. This method, while successful in implementing spatially varying regularization, is not suitable for end-to-end registration networks because it requires initial momentum for network input. Additionally, the number and standard deviations of the Gaussian kernels, as well as the weighting parameters for OMT and TV losses, are hyperparameters that need extensive training cycles to optimize manually.

\section{Proposed Method}
\begin{figure}[t]
\begin{center}
\includegraphics[width=0.75\textwidth]{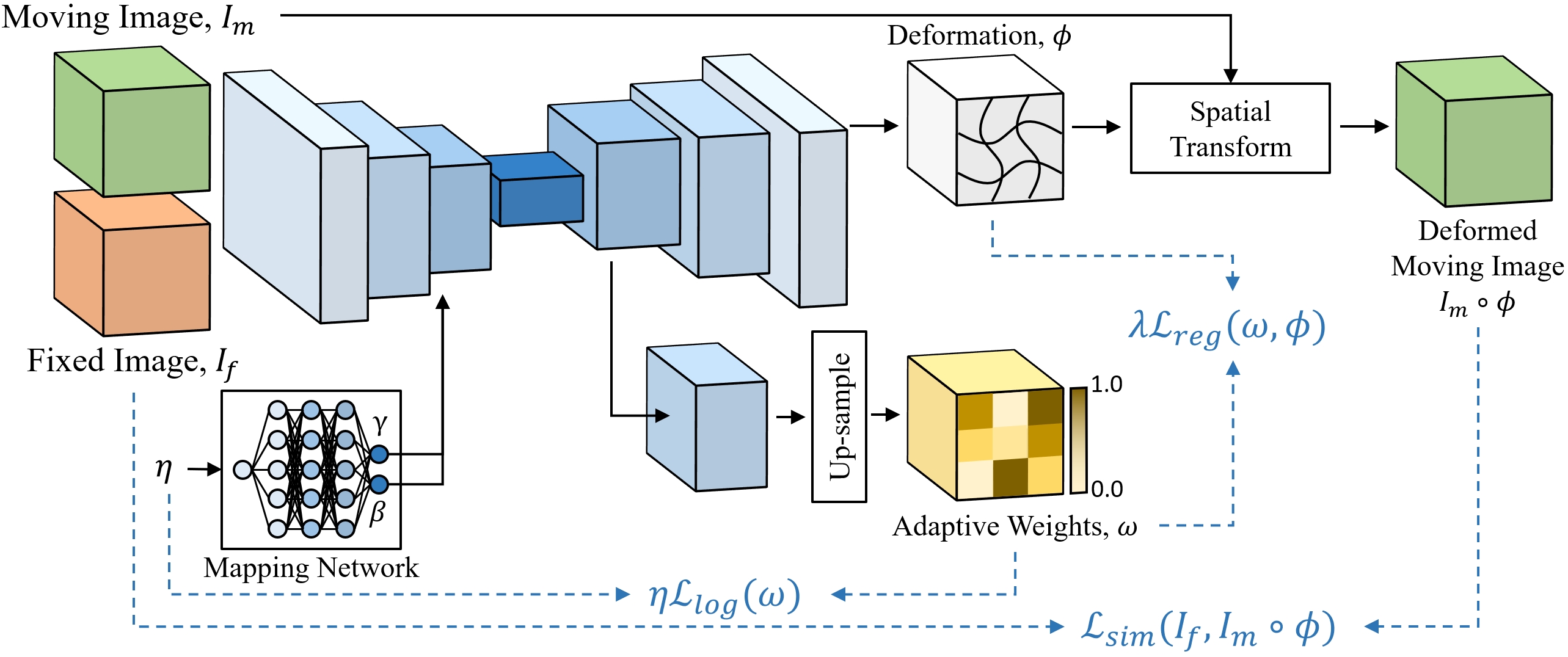}
\end{center}
   \caption{The overall framework of the proposed method. }
\label{fig:framework}
\end{figure}

\noindent{\textbf{Weighted Diffusion Regularizer. }}
We aim to develop a spatially-varying deformation regularizer for end-to-end learning of deformable image registration, rather than a spatially invariant regularization for the entire deformation as used in almost all existing learning-based registration methods~\cite{balakrishnan2019voxelmorph,kim2021cyclemorph,chen2022transmorph,mok2021conditional}. As shown in Fig. \ref{fig:framework}, we used a neural network to take in the moving and fixed images, $I_m$ and $I_f$, which are defined over a 3D spatial domain $I_m, I_f\in\mathbb{R}^{H\times W\times D}$. The network outputs a deformation field $\phi\in\mathbb{R}^{3\times H\times W\times D}$ that warps $I_m$ to $I_f$, as well as a locally adaptive weight volume $\omega\in\mathbb{R}^{H\times W\times D}$. We then used this weight volume to apply spatially-varying levels of regularization to different voxels through a \textit{weighted diffusion regularizer}:
\begin{equation}
    \label{eqn:wt_diff}
    \lambda\mathcal{L}_{reg}(\omega, \phi)=\lambda\sum_{\mathbf{p}\in\Omega}\omega(\mathbf{p})\Vert\nabla\pmb{u}(\mathbf{p})\Vert^2,
\end{equation}
where $\lambda$ controls the strength of regularization, $\omega(\mathbf{p})\in[0, 1]$ is the weight corresponds to voxel location $\mathbf{p}$, and $\pmb{u}$ denotes the displacement field. Since $\lambda$ is related to the variance of the Normal distribution (as seen in section \ref{sec:background}), a larger value of $\omega(\mathbf{p})$ would impose a stronger Gaussian smoothing for the voxel location $\mathbf{p}$, with $\lambda$ being the highest possible strength (i.e., the largest variance of the kernel). In fact, this weighted diffusion regularizer can be thought of as applying spatially-changing multi-Gaussian kernels to deformation, as used in \cite{niethammer2019metric} (section \ref{sec:background}). This is because the combination of multiple Gaussians is in fact a Gaussian. However, we argue that the proposed regularizer is more suitable for end-to-end training than the method in \cite{niethammer2019metric} as it is integrated into a loss function and does not require pre-determining the number and variances of the Gaussians.

In the decoder, as shown in Fig. \ref{fig:framework}, an additional convolution block consisting of three convolutional layers was applied to the 1/4 resolution branch to generate the adaptive weight volume $\omega$. The output was then passed through a sigmoid function to constrain the weight values to a range of $[0,1]$. The final weight volume was then up-sampled to match the resolution of the deformation field. By generating the weight volume at a lower resolution, we were able to reduce the computational workload and introduce some spatial smoothness in the weight volume, which aligns with the physical interpretation of these weights.

\noindent{\textbf{Log Loss. }}
Eqn. \ref{eqn:wt_diff} by itself is not enough to prevent the network from reaching trivial solutions, such as always producing an output of zero for $\omega(\mathbf{p})$. Additionally, it would be desirable to favor smoother deformations when possible. We therefore introduced a penalty to discourage the generation of small $\omega$ values. Specifically, we designed a \textit{log loss} based on binary cross-entropy, but it can be simplified by using predefined target values:
\begin{equation}
    \label{eqn:log_loss}
    \eta\mathcal{L}^*_{log}(\omega)=-\eta\sum_{\mathbf{p}\in\Omega}\bigg(1\cdot\log(\hat{\omega}(\mathbf{p}))+0\cdot\log(1-\hat{\omega}(\mathbf{p}))\bigg)=-\eta\sum_{\mathbf{p}\in\Omega}\log(\hat{\omega}(\mathbf{p})),
\end{equation}
where $\eta$ controls the strength of the log loss, $\hat{\omega}(\mathbf{p})=\text{clamp}_{\epsilon, 1}(\omega(\mathbf{p}))$ is the weight that is clamped to the range of $[\epsilon, 1]$ in order to prevent the error from occurring due to $\log(0)$, and $\epsilon$ is set to $1e^{-4}$ for all experiments. This loss becomes zero if $\hat{\omega}(\mathbf{p})=1$ and has its largest value when $\hat{\omega}(\mathbf{p})=\epsilon$. Subsequently, we normalized the loss by dividing it by its maximum, yielding $\eta\mathcal{L}_{log}(\omega)=\eta\frac{\mathcal{L}_{log}^*(\omega)}{\log(\epsilon)}$.

With the loss functions introduced in previous sections, the overall loss function for training the registration network is expressed as follows:
\begin{equation}
    \label{eqn:loss}
    \mathcal{L}(I_f, I_m, \phi, \omega)=\mathcal{L}_{sim}(I_f, I_m)+\lambda\mathcal{L}_{reg}(\phi, \omega)+\eta\mathcal{L}_{log}(\omega),
\end{equation}
In contrast to conventional learning-based methods that only use the first two terms for network training, the proposed method incorporates an additional log loss term. While this may appear to make the tuning of the method more complicated due to the additional hyperparameter, the use of $\lambda$ as an upper bound for regularization strength allows for easy tuning of the loss function in practice. By setting $\lambda$ to a fixed and large value, the network will adaptively adjust the weights (i.e., $\omega$) based on the chosen $\eta$, eliminating the need for additional tuning of $\lambda$.
\begin{figure}[t]
\begin{center}
\includegraphics[width=0.85\textwidth]{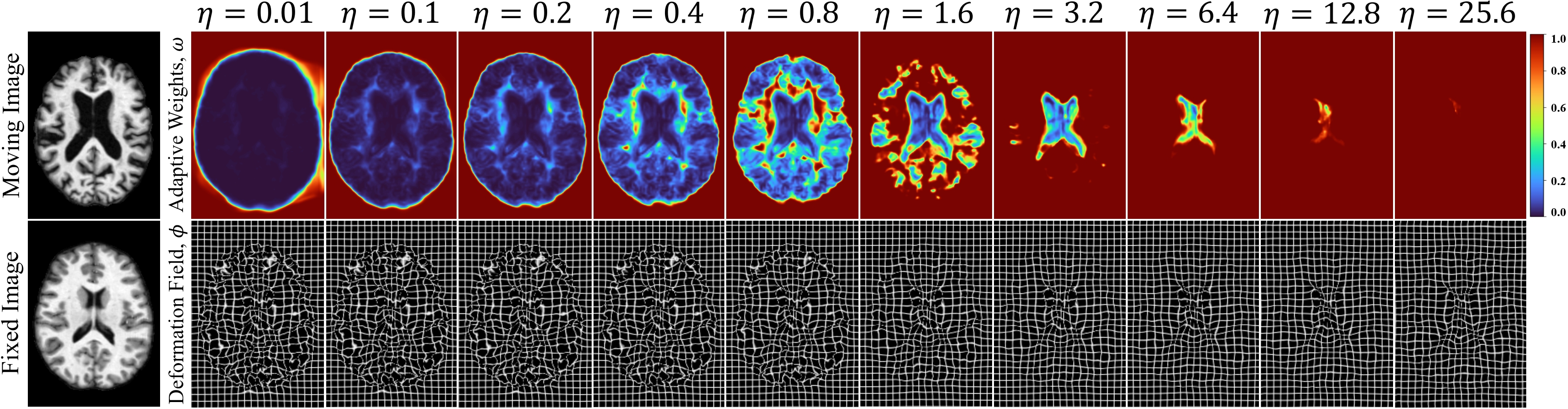}
\end{center}
   \caption{The figure illustrates the impact of varying the weight, $\eta$, of the log loss on the smoothness of the deformation fields. }
\label{fig:eta_1}
\end{figure}

\noindent{\textbf{Conditional Transformer. }}
We went one step further to adopt the hyperparameter conditioning method proposed in \cite{mok2021conditional,hoopes2021hypermorph}, in which the hyperparameter of the regularizer is conditioned into the network by sampling different values during training. The hyperparameter is then self-tuned through inference by testing different values and selecting the one that yields the highest Dice score on the validation dataset. Here, we conditioned the weight of the log loss (\emph{i.e.}, $\eta$) into the network architecture, as shown in Fig. \ref{fig:framework}. This allows us to tune the value of $\eta$ through a \textit{single} training of the registration network. We used \texttt{TM-TVF}~\cite{chen2022transmorph, chen2022unsupervised}, a Transformer-based network that has shown promising results on various registration tasks, as the backbone network. We built upon the \textit{conditional instance normalization} (CIN) proposed in \cite{mok2021conditional}, where the control of regularization strength is learned by normalizing and shifting feature map statistics by two affine parameters, $\gamma$ and $\beta$, obtained from a mapping network given $\eta$. Here, we extended CIN to \textit{conditional layer normalization} (CLN) for its use in the Transformer encoder of \texttt{TM-TVF}. CLN is expressed as follows:
\begin{equation}
    \pmb{h}'_i=\gamma_i\frac{\pmb{h}_i-\mu(\pmb{h}_i)}{\sigma(\pmb{h}_i)} + \beta_i,
\end{equation}
where $\pmb{h}_i$ denotes the features of the $i^{th}$ hidden layer, with $\mu$ and $\sigma$ being the mean and standard deviation of all hidden units within that layer~\cite{ba2016layer}. $\gamma_i$ and $\beta_i$ are produced by applying the MLPs to $\eta$ (\emph{i.e.}, $\gamma_i,\beta_i=\text{MLP}_i^{\gamma,\beta}(\eta))$). The MLP architecture used here was identical to that in \cite{mok2021conditional}, with a latent space dimension of 64. In \texttt{TM-TVF}, the original layer norm and instance norm found in the Transformer encoder and ConvNet decoder were replaced with CLN and CIN, respectively.

For the rest of this paper, the proposed method is denoted as \texttt{TM-SPR} to emphasize the usage of the spatially-varying regularizer (SPR). Moreover, we employed the \textit{scaling-and-squaring} (SS) approach~\cite{arsigny2006log, dalca2019unsupervised} to ensure diffeomorphic registration. The resulting model is denoted as \texttt{TM-SPR}$_{\text{\textit{diff}}}$.

\section{Experiments}
\noindent{\textbf{Dataset and Pre-processing. }} The proposed method was evaluated on inter-patient and atlas-to-patient registration tasks. The OASIS dataset~\cite{marcus2007open}, obtained from the Learn2Reg challenge~\cite{hering2022learn2reg}, was used for inter-patient registration. Additionally, the IXI dataset\footnote{https://brain-development.org/ixi-dataset/} obtained from \cite{chen2022transmorph} was employed for the atlas-to-patient registration task. The OASIS dataset encompasses 413 T1-weighted brain MRI images, with 394 volumes designated for training and 19 volumes allocated for validation. The IXI dataset, on the other hand, comprises 576 T1-weighted brain MRI images, with a distribution of 403 volumes for training, 58 volumes for validation, and 115 volumes for testing. Additionally, for the atlas-to-patient registration task, a moving image, which was a brain atlas image obtained from \cite{kim2021cyclemorph}, was used. The pre-processed image volumes were all subsequently cropped to the dimensions of $160 \times 192 \times 224$. Anatomical label maps, including over 30 anatomical structures, were generated to evaluate registration performances.

\begin{figure}[t]
\begin{center}
\includegraphics[width=0.85\textwidth]{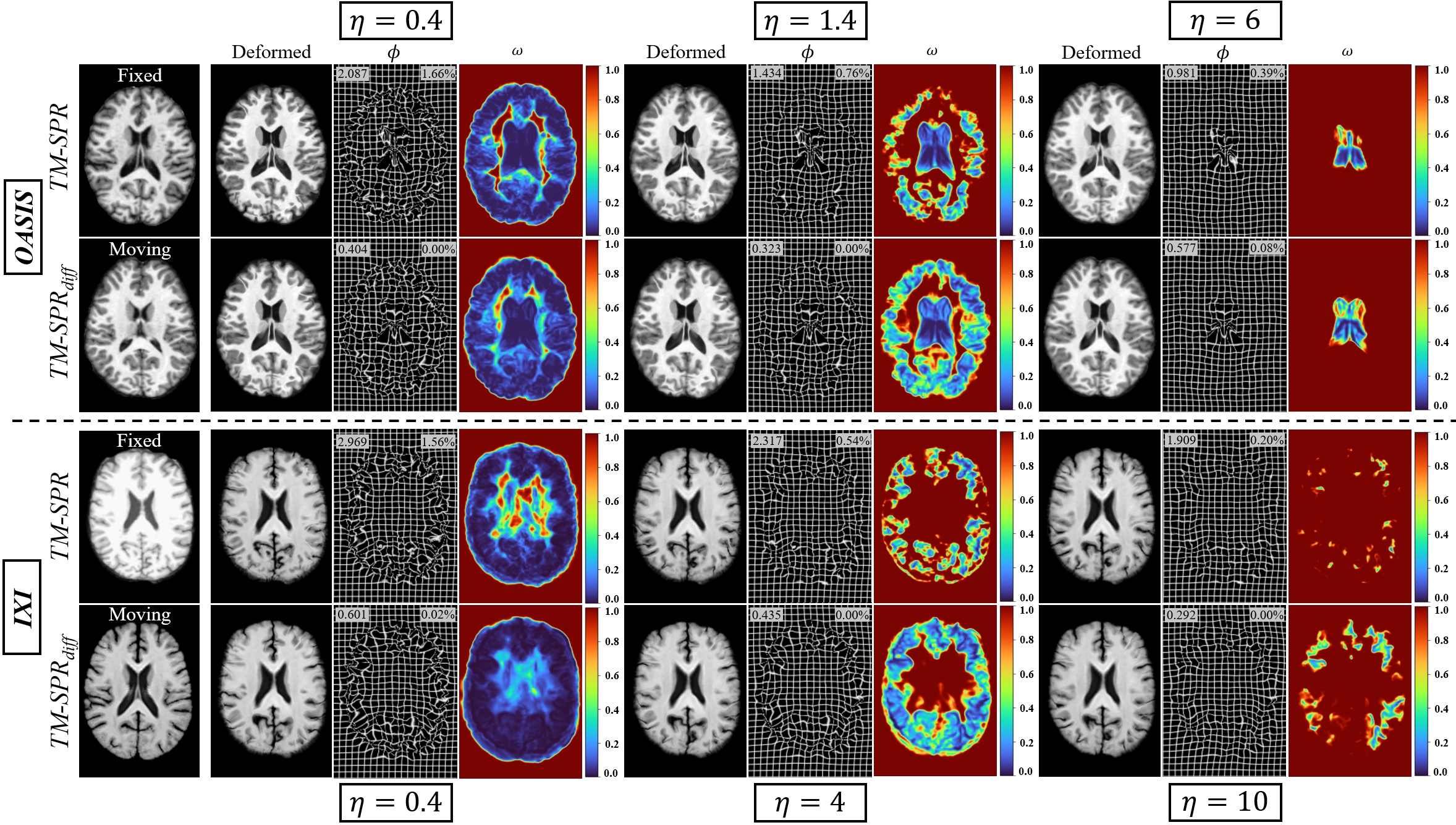}
\end{center}
   \caption{Qualitative results of the proposed method on the OASIS and IXI datasets. The values in the deformation fields quantify the deformation smoothness, with the left value denoting SDlogJ and the right value denoting \%NDV.}
\label{fig:eta_2}
\end{figure}

\noindent{\textbf{Implementation Details. }} The models were trained for 500 epochs using the Adam optimizer~\cite{kingma2014adam} and a batch-size of 1. The learning rate was set to 0.0001. The hyperparameter $\lambda$ for the weighted diffusion regularizer was consistently set to 5 across all experiments. During training, $\eta$ was uniformly sampled within the range of $[0,2]$ for OASIS and $[0,4]$ for IXI, and the values were subsequently normalized to $[0,1]$ before being fed into the network. The number of time steps in \texttt{TM-TVF}~\cite{chen2022unsupervised} was set to 7. In all experiments, NCC was used as $\mathcal{L}_{sim}$. Additionally, for OASIS, the Dice loss was used to take advantage of the supervision provided by anatomical label maps, whereas it was not used for IXI. 
%The PyTorch framework was employed for the implementation of the models, and training was conducted on an NVIDIA RTX3090 GPU. 

\noindent{\textbf{Evaluation Metrics. }} To evaluate registration performance, we measured the overlap of the anatomical label maps using Dice. Additionally, we computed the Hausdorff distance (HdD) and the standard deviation of the Jacobian determinant (SDlogJ) for the OASIS dataset to align with the leaderboard of the Learn2Reg challenge. Whereas, for the IXI dataset, we employed the percentage of all non-positive Jacobian determinant (\%$\vert J\vert\leq0$) and the non-diffeomorphic volume (\%NDV), both proposed in \cite{liu2022finite}, to assess deformation invertibility as they are more accurate measures under the finite-difference approximation.

\noindent{\textbf{Results and Discussion. }}
Fig. \ref{fig:eta_1} shows the impact of $\eta$ on the adaptive weight volume and the smoothness of the deformation field. The weight volume allocates different yet smooth regularization strengths to different regions in the image. As $\eta$ gradually increases, the values in weight volumes shift towards 1 for all regions, progressively resulting in a smoother deformation globally. Fig. \ref{fig:eta_2} shows the additional qualitative examples of the proposed method on the two datasets. Note that for the OASIS dataset, the ventricle regions were assigned smaller values of $\omega$, indicating weaker regularization. Conversely, for the IXI dataset, the ventricle regions were assigned larger values of $\omega$ that were closer to 1. This difference may be attributed to the use of the Dice loss for the OASIS dataset, which places strong and explicit constraints on anatomical overlaps. Moreover, it is evident that the proposed adaptive regularization approach allows for smaller values of $\omega$ under the same value of $\eta$ when the SS approach is used. These visual results serve as evidence of the adaptiveness and robustness of the proposed method in accommodating different loss and network configurations.

\begin{table}[t]
\centering
\fontsize{5.5}{7}\selectfont
    \begin{tabular}{ c | c c c m{0.0001\textwidth} c | c c c}
 \Xhline{1pt}
 \multicolumn{4}{c}{\textbf{OASIS}}& &\multicolumn{4}{c}{\textbf{IXI}}\\
 \cline{1-4}\cline{6-9}
 Method & Dice$\uparrow$ & HdD95$\downarrow$ & SDlogJ$\downarrow$&& Method & Dice$\uparrow$ & \%$\vert J\vert\leq0\downarrow$ & \%NDV$\downarrow$ \\
 \cline{1-4}\cline{6-9}
 ConvexAdam~\cite{siebert2021fast} & 0.846$\pm$0.016  & 1.500$\pm$0.304& 0.067$\pm$0.005 && VoxelMorph~\cite{balakrishnan2019voxelmorph} & 0.732$\pm$0.123&6.26\%&1.04\%\\
\cline{1-4}\cline{6-9}
 LapIRN~\cite{mok2021conditional} & 0.861$\pm$0.015  & 1.514$\pm$0.337& 0.072$\pm$0.007 && CycleMorph~\cite{kim2021cyclemorph} & 0.737$\pm$0.123&6.38\%&1.15\%\\
 \cline{1-4}\cline{6-9}
 TransMorph~\cite{chen2022transmorph} & 0.862$\pm$0.014  & \textit{1.431$\pm$0.282}& 0.128$\pm$0.021 && TransMorph~\cite{chen2022transmorph} & 0.754$\pm$0.124&5.65\% &0.90\%\\
 \cline{1-4}\cline{6-9}
  TM-TVF~\cite{chen2022unsupervised} & \textit{0.869$\pm$0.014}& \textbf{1.396$\pm$0.295} & 0.094$\pm$0.018 && TM-TVF~\cite{chen2022unsupervised} & 0.756$\pm$0.122 & 2.05\% & 0.36\%\\
 \Xhline{1pt}%\cline{1-4}\cline{6-9}
 TM-SPR & \textbf{0.870$\pm$0.017}  &1.483$\pm$0.397& 0.234$\pm$0.05&& TM-SPR & \textbf{0.769$\pm$0.123} & 2.02\% & 0.52\%\\
 \cline{1-4}\cline{6-9}
 TM-SPR$_{\text{\textit{diff}}}$ & 0.852$\pm$0.014 & 1.597$\pm$0.345& 0.082$\pm$0.03 && TM-SPR$_{\text{\textit{diff}}}$ & \textit{0.762$\pm$0.123} & 0.03\% & 0\%\\
 \Xhline{1pt}
\end{tabular}
\caption{Quantitative results for inter-patient (OASIS) and atlas-to-patient (IXI) registration tasks. Note that part of the OASIS results was obtained from Learn2Reg leaderboard~\cite{hering2022learn2reg}.}\label{tab:oasis_IXI}
\end{table}

\begin{figure}[t]
\begin{center}
\includegraphics[width=0.7\textwidth]{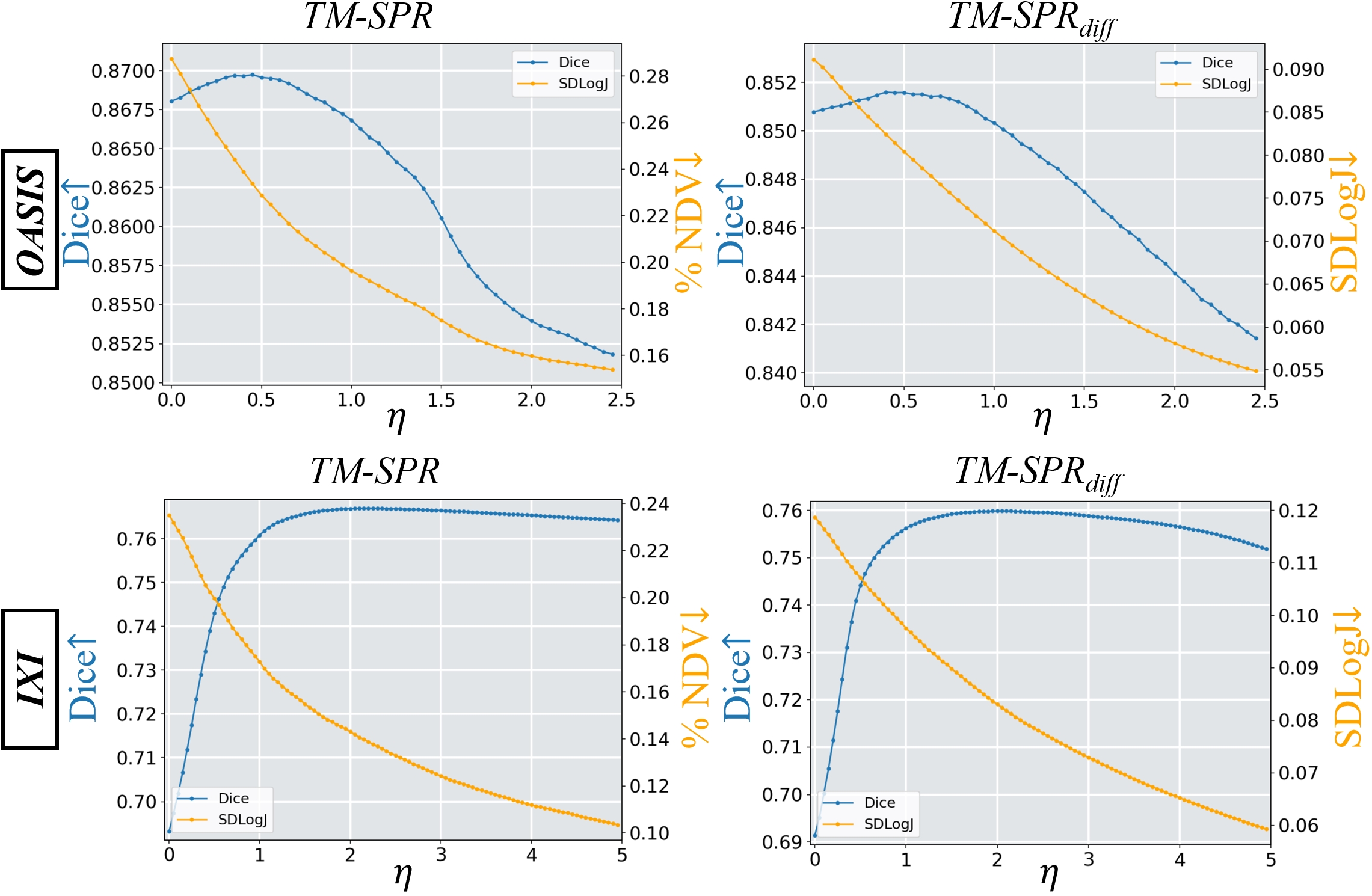}
\end{center}
   \caption{The plots show the relationship between the weight $\eta$ and the registration performance measured by Dice, as well as the smoothness of deformation represented by \%NDV and SDLogJ.}
\label{fig:reg_performance}
\end{figure}

A grid search was conducted to find the optimal $\eta$ values for the validation datasets of the two registration tasks, with a step size of 0.05, as illustrated in Fig. \ref{fig:reg_performance}. The resulting $\eta$ values that produced the highest Dice scores were 0.45 and 0.4 for \texttt{TM-SPR} and \texttt{TM-SPR}$_{\text{\textit{diff}}}$ on the OASIS dataset, and 2.2 and 2.0 on the IXI dataset. The proposed method was evaluated against several state-of-the-art methods and the quantitative results are shown in Table \ref{tab:oasis_IXI}. On the OASIS dataset, the performance of \texttt{TM-SPR} is comparable to its base method \texttt{TM-TVF} with reduced smoothness. This is likely due to the use of the Dice loss during training, which causes the network to produce lower regularization strength to prioritize greater anatomical matching. However, the proposed method demonstrated significant advantages when anatomical label maps were not used during training. On the IXI dataset, despite having a nearly identical architecture to \texttt{TM-TVF}, \texttt{TM-SPR} demonstrated significantly better performance in Dice score than all other comparative methods. This improvement was statistically significant with $p\ll0.0001$ from a paired t-test when compared to \texttt{TM-TVF}, while maintaining the same level of deformation smoothness as \texttt{TM-TVF}. It is noteworthy that \texttt{TM-SPR}$_{\text{\textit{diff}}}$ produced the second-best Dice score and imposed a diffeomorphic registration with almost no folded voxels.

\section{Conclusion}
In this study, we introduced a simple yet effective framework for end-to-end learning of spatially-varying regularization for DNN-based image registration. Unlike most DNN-based methods that apply spatially invariant regularization, our method learns to generate an adaptive weight volume that assigns individual weights to voxels for applying spatially-varying levels of regularization. A novel penalty term was proposed to compel the network to impose stronger regularization when possible. The weighting parameter for the penalty term was conditioned into the network to facilitate easy hyperparameter tuning. Qualitative and quantitative results shown in this study demonstrate the effectiveness of the proposed framework for medical image registration.
\bibliographystyle{splncs04}
\bibliography{mybibliography}

\clearpage
\section*{Appendix}
\appendix
\section{Prior Distribution under Gaussian Assumption}
Let the distribution over $\pmb{u}$ be a multivariate Normal distribution, $p(\pmb{u}) \propto \mathcal{N}(\pmb{u};\barbelow{\mathbf{0}},\mathbf{\Sigma})$. The logarithm of $p(\pmb{u})$ can be derived as follows:
\begin{equation}
\begin{split}
   \log p(\pmb{u})&=-\frac{1}{2}\log(\mathbf{\Sigma})-\pmb{u}^\intercal\mathbf{\Sigma}^{-1}\pmb{u} + \text{const.}\\
   &=\frac{1}{2}\log(\mathbf{\Sigma}^{-1})-\pmb{u}^\intercal\mathbf{\Sigma}^{-1}\pmb{u} + \text{const.}\\
   &=\frac{1}{2}\log(\mathbf{\Lambda}_{\pmb{u}})-\pmb{u}^\intercal\mathbf{\Lambda}_{\pmb{u}}\pmb{u} + \text{const.}\\
   &=-\pmb{u}^\intercal\mathbf{\Lambda}_{\pmb{u}}\pmb{u} + \text{const.} \approx -\lambda\sum_{\mathbf{p}\in\Omega}\Vert\nabla \pmb{u}(\mathbf{p})\Vert^2,
\end{split}  
\end{equation}
where $\mathbf{\Sigma}^{-1}=\mathbf{\Lambda}_{\pmb{u}}=\lambda \mathbf{L}$, and $\log\vert\mathbf{\Lambda}_{\pmb{u}}\vert$ reduces to a constant. Here, $\mathbf{L}$ denotes the Laplacian of a neighborhood graph defined on the image grid, and $\lambda$ is a parameter that controls the scale of the displacement.

\section{ConvNet Block for Adaptive Weight Volume}
\begin{figure}[!htb]
\begin{center}
\includegraphics[width=0.6\textwidth]{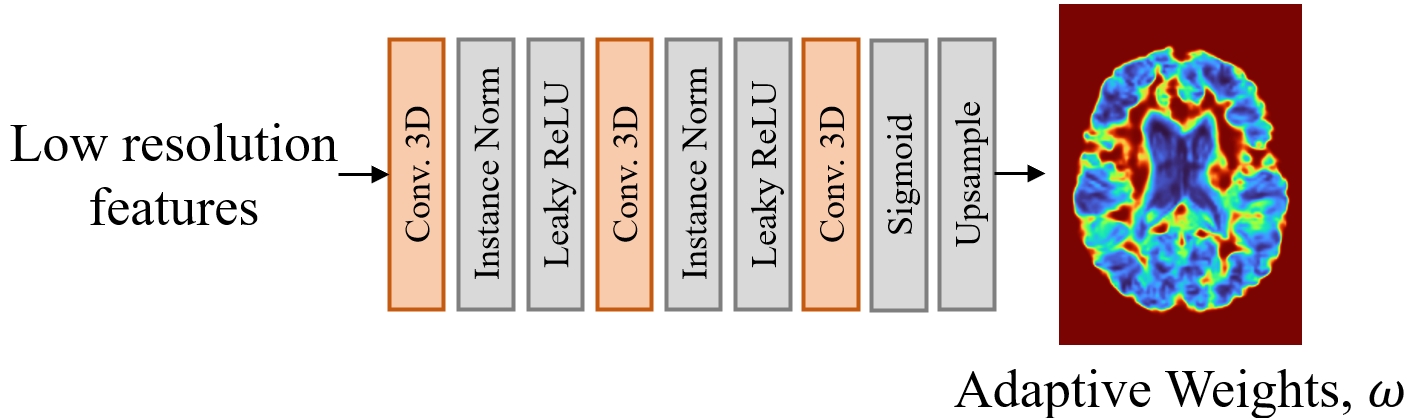}
\end{center}
   \caption{The schematic of the ConvNet block that produces the adaptive weight volume, $\omega$.}
\label{fig:ConvNet_for_w}
\end{figure}

\clearpage
\section{Data Preprocessing}
The pre-processing steps for both the OASIS and IXI datasets were similar, where FreeSurfer~\cite{fischl2012freesurfer} was employed to carry out standard procedures for structural brain MRI, including skull stripping, resampling, and affine transformation. 

\section{Additional Quantitative Results}
\begin{figure}[!htb]
\begin{center}
\includegraphics[width=1.\textwidth]{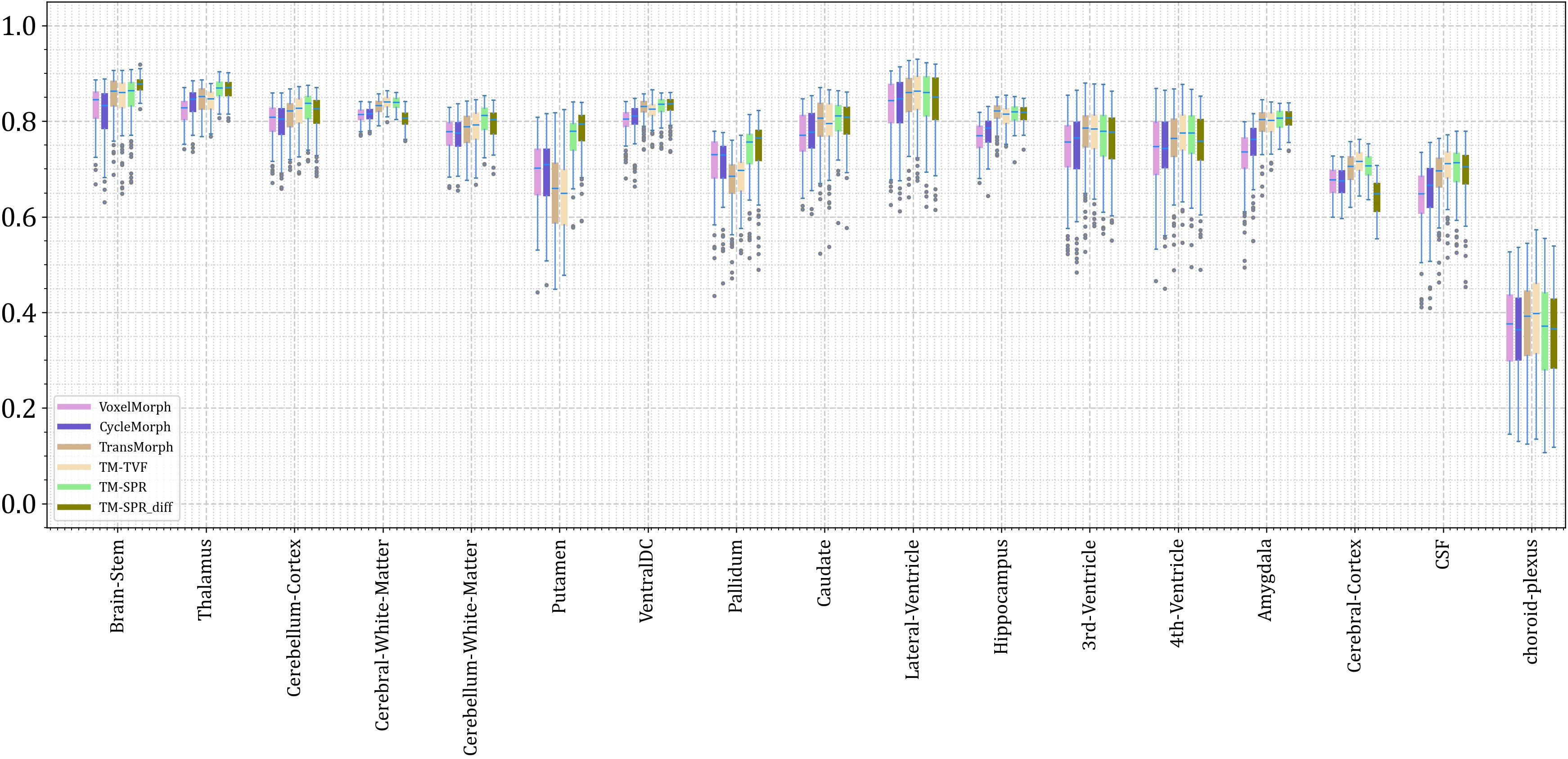}
\end{center}
   \caption{Quantitative results of atlas-to-patient registration on the IXI dataset.}
\label{fig:Albers_quant}
\end{figure}

\end{document}